\documentclass[sigconf, screen]{acmart}
\usepackage{braket}
\usepackage{tabularx}
\usepackage{multirow}
\usepackage{listings}
\usepackage[super]{nth}

\AtBeginDocument{%
  \providecommand\BibTeX{{%
    \normalfont B\kern-0.5em{\scshape i\kern-0.25em b}\kern-0.8em\TeX}}}

\acmYear{2019}\copyrightyear{2019}
\setcopyright{acmcopyright}
\acmConference[ISCA '19]{The 46th Annual International Symposium on Computer Architecture}{June 22--26, 2019}{Phoenix, AZ, USA}
\acmBooktitle{The 46th Annual International Symposium on Computer Architecture (ISCA '19), June 22--26, 2019, Phoenix, AZ, USA}
\acmPrice{15.00}
\acmDOI{10.1145/3307650.3322213}
\acmISBN{978-1-4503-6669-4/19/06}

\begin{document}

\title[Statistical Assertions for Validating Patterns and Finding Bugs in Quantum Programs]{Statistical Assertions for Validating Patterns and\\ Finding Bugs in Quantum Programs}

\author{Yipeng Huang}
\email{yipeng@cs.princeton.edu}
\orcid{0000-0003-3171-6901}
\affiliation{%
  \institution{Princeton University}
%   \streetaddress{35 Olden Street}
%   \city{Princeton}
%   \state{New Jersey}
%   \postcode{08540-5233}
}

\author{Margaret Martonosi}
\email{mrm@princeton.edu}
\affiliation{%
  \institution{Princeton University}
%   \streetaddress{35 Olden Street}
%   \city{Princeton}
%   \state{New Jersey}
%   \postcode{08540-5233}
}

% \renewcommand{\shortauthors}{Huang and Martonosi}

\begin{abstract}
In support of the growing interest in quantum computing experimentation,
programmers need new tools to write quantum algorithms as program code.
Compared to debugging classical programs, debugging quantum programs is difficult because programmers have limited ability to probe the internal states of quantum programs;
those states are difficult to interpret even when observations exist;
and programmers do not yet have guidelines for what to check for when building quantum programs.
In this work, we present quantum program assertions based on statistical tests on classical observations.
These allow programmers to decide if a quantum program state matches its expected value in one of classical, superposition, or entangled types of states.
We extend an existing quantum programming language with the ability to specify quantum assertions,
which our tool then checks in a quantum program simulator.
We use these assertions to debug three benchmark quantum programs in factoring, search, and chemistry.
We share what types of bugs are possible,
and lay out a strategy for using quantum programming patterns to place assertions and prevent bugs.
\end{abstract}




\maketitle

% Simulating a sectioning command by setting the first word or words of
% a paragraph in boldface or italicized text is {\bfseries not allowed.}

% Table captions are placed {\itshape above} the table.

% \begin{table}
%   \caption{Frequency of Special Characters}
%   \label{tab:freq}
%   \begin{tabular}{ccl}
%     \toprule
%     Non-English or Math&Frequency&Comments\\
%     \midrule
%     \O & 1 in 1,000& For Swedish names\\
%     $\pi$ & 1 in 5& Common in math\\
%     \$ & 4 in 5 & Used in business\\
%     $\Psi^2_1$ & 1 in 40,000& Unexplained usage\\
%   \bottomrule
% \end{tabular}
% \end{table}

% \begin{table*}
%   \caption{Some Typical Commands}
%   \label{tab:commands}
%   \begin{tabular}{ccl}
%     \toprule
%     Command &A Number & Comments\\
%     \midrule
%     \texttt{{\char'134}author} & 100& Author \\
%     \texttt{{\char'134}table}& 300 & For tables\\
%     \texttt{{\char'134}table*}& 400& For wider tables\\
%     \bottomrule
%   \end{tabular}
% \end{table*}

% \begin{figure}[h]
%   \centering
%   \includegraphics[width=\linewidth]{sample-franklin}
%   \caption{1907 Franklin Model D roadster. Photograph by Harris \&
%     Ewing, Inc. [Public domain], via Wikimedia
%     Commons. (\url{https://goo.gl/VLCRBB}).}
%   \Description{The 1907 Franklin Model D roadster.}
% \end{figure}

% Figure captions are placed {\itshape below} the figure.

\begin{figure}[t]
\centering
\includegraphics[width=\columnwidth]{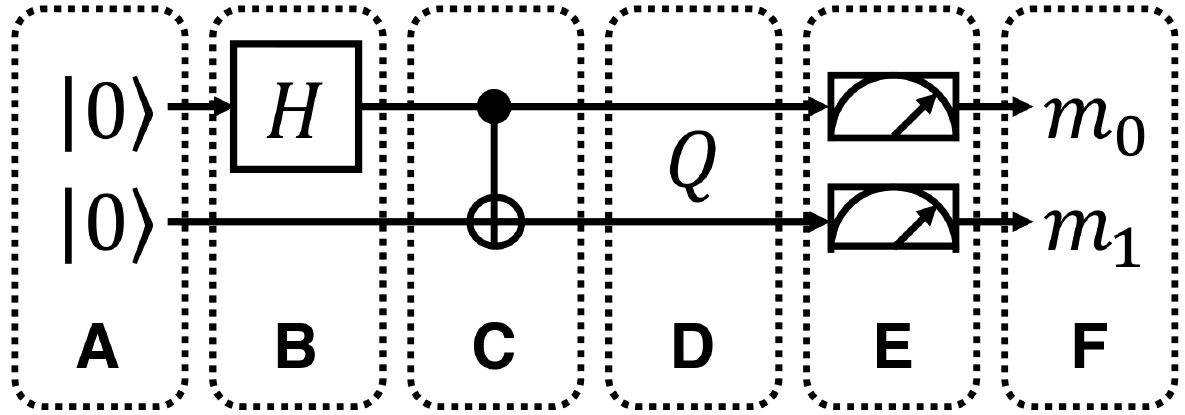}
\Description[A Bell state creation quantum program]{A schematic showing the evolution of two qubits, through seven stages labeled A through F.}
\caption{A Bell state creation quantum program.
The sequence of operations in time flows left to right.
A classical state consisting of two qubits (A) is manipulated into a superposition state by a quantum operation (B).
A controlled-NOT gate (C) then induces entanglement between the two qubits to create an entangled state $Q$,
which notably can no longer be factored into two separate pieces of information (D).
Measurement of both of the qubits collapses the quantum state (E).
Because the the qubits were entangled when they were measured, the measurement results $m_0$ and $m_1$ are correlated (F).
Statistical tests on these measurements aid programmers in implementing and debugging quantum programs.
}
\label{fig:bell}
\end{figure}

\section{Introduction}

% Motivation
Quantum computing is reaching an inflection point~\cite{Preskill2018computinginnisqera, NAP25196}.
After years of work on low-level quantum computing (QC) devices, small but viable QC prototypes are now available to run programs.
These QC prototypes are increasing in size,
with much research attention being placed on improving their reliability and increasing the counts of qubits (quantum bits),
the fundamental building block for QC~\cite{ladd_jelezko, experimental_comparison, tannu_asplos, murali_asplos}.
These advancements in QC hardware may soon lead to a demonstration of a quantum algorithm running on QC hardware that exceeds the performance of a classical computer system~\cite{Harrow,boixo2018supremacy}.
Such a demonstration would move the world closer to a new era of computing where QC systems solve problems in chemistry~\cite{qchem,qchem_nsf}, optimization~\cite{farhi_adiabatic,farhi_qaoa,pakin_asplos}, and even cryptography~\cite{shor} that are currently intractable with classical computers.

With small-scale machines available to run real code,
a natural challenge lies in bridging the QC architectural gap between algorithms and hardware.
One aspect of that gap is in creating correct programs to run on quantum computers~\cite{nature, software_methodology}.
Until recently, QC algorithms existed only in the form of abstract specifications and equations,
and were rarely programmed for actual execution or simulation,
and therefore relatively little QC debugging has ever occurred.
Furthermore, QC debugging faces challenges beyond that of classical computing.
In particular, typical debugging approaches based on printing out variable values during program execution do not easily apply to QC programs,
because program states in QC collapse to classical values when observed.
Second, while programmers have more freedom to observe full program states in QC simulations on classical computers, the massive state spaces of QC executions limits this approach to small programs.
Finally, even when limited simulations are tractable, it can be difficult to interpret the simulation results.

While the problem of debugging and validating quantum programs has been extensively identified as a major barrier to useful quantum computation~\cite{nature,software_methodology,quipper,svore_quantum_future,design_automation,qwire,NAP25196},
little has been said about what actually constitutes a quantum program bug.
Similarly limited detail has been shared about the inside story of translating QC algorithms in to working QC programs,
even though the field is now making rapid progress in writing open source QC program benchmarks across several quantum programming languages~\cite{comparison,quipper,scaffcc,Steiger2018projectqopensource,q_sharp,openfermion}.

This work shares the detailed process of debugging quantum programs,
with the help of a proposed set of quantum program assertions and breakpoints based on statistical tests.
Using ensembles of classical observations taken from the intermediate state of quantum programs,
these statistical tests are able to decide if the program state belongs to one of three important classes of quantum states: \emph{classical}, \emph{superposition}, and \emph{entangled}.
With this information,
programmers can determine if the execution of the quantum program is valid up to each breakpoint.
If the program state is invalid, the assertions guide the programmer in finding the bug inside subroutines and in program code up to that point.

For three quantum program benchmarks in factoring integers, database search, and quantum chemistry,
we describe what kinds of bugs occurred in the process of bringing up the programs from unit tests to integration testing.
We categorize the bugs according to where in the structure of quantum programs they may arise,
and we lay out a strategy for placing statistical assertions that effectively catches them.

The rest of this paper is organized as follows:
Section~\ref{sec:statics} provides relevant background for quantum programs and debugging.
Section~\ref{sec:methodology} details our statistical assertions and simulation framework for debugging quantum programs,
which is then used in Section~\ref{sec:shors} for building and debugging an integer factorization quantum program.
Section~\ref{sec:algorithms} evaluates the use of the assertions in two additional case studies.
Section~\ref{sec:related} discusses related approaches to writing correct quantum programs.

% Pull from presentation slides 
% Possibly pull tutorial picture from plateau slides into section 2

\section{Background on quantum states and quantum programs}
\label{sec:statics}
% Tutorial on quantum computing
% Why is debugging challenging
First, we review the principles of quantum computing~\cite{Kaye:2007:IQC:1206629, mermin2007quantum, metodi, nielsen_chuang},
in order to understand how debugging quantum programs is different from and more challenging than classical debugging.

\subsection{Qubits, superpositions, and entanglement}

The basic unit of information in QC is the \emph{qubit},
which can take on values of $\ket{0}$ and $\ket{1}$ like bits in classical computing,
but unlike classical bits, qubits can also be in a probabilistic \emph{superposition} between the two values.
% quantum superpositions (induced by Hadamard gate)
% quantum variables' ability to be in superposition underlies power of quantum computing
Quantum computers can also \emph{measure} the value of a qubit,
forcing it to collapse out of superposition into a classical value such as `0' or `1'.
Measurement disturbs the values of variables in a quantum computer.
So unlike the case in classical computing,
programmers cannot easily pause execution and observe the values of qubits as a quantum program runs.
As a result of this limited visibility into program state,
programmers must carefully choose what to measure and test when they debug quantum programs.
% no cloning

% In an atom analogy
% we have |0>, the ground state
% |1> excited state
% In the ground state we can shine light of a certain wavelength (energy) to excite it
% |y> = a|0> + b|1>, where a,b belong to C and $|a|^2 + |b|^2 = 1$
% The hypotenuse length is always 1
% Global phase has no measurable effect, at least at the level of a single isolated qubit

Aside from qubits being in superposition states,
the other feature of data in QC is \emph{entanglement}.
% Tensor products
For example, when the states of two qubits are entangled,
the combined state of the two-qubit system must be viewed as a superposition of a larger set of elementary states $\ket{00}$, $\ket{01}$, $\ket{10}$, and $\ket{11}$.
An entangled state cannot be factored into independent pieces of information,
and can no longer be viewed as a simple concatenation of two qubits.
Likewise, a three-qubit system has potential superpositions of eight states, and so on.
For this reason, as more qubits come into play in a quantum computer,
the number of states that data can be in grows exponentially.
This exponential growth of possible values due to superpositions and entanglement underlies the power of QC.

On the flip side, the huge state spaces involved in QC limits programmers' ability to use classical computers to simulate and debug QC programs.
Na\"ive simulation of a 50-qubit quantum computer, for example, needs $2^{50}$ or roughly one quadrillion floating point numbers just to store the program state at any instant~\cite{haner_petabyte}.
While more advanced techniques can decrease the memory requirement for simulating circuits~\cite{zulehner,thornton,tensor,viamontes,wu_compression},
interactive programming and simulating quantum programs on a workstation is still limited to 20 to 30 qubits.
For this reason, testing and debugging QC programs in simulation is only possible for toy-sized programs.

% Finally, physical computers limited in time, size, and reliability
\subsection[Quantum computer operations, programs, and a taxonomy for bugs]{Quantum computer operations,\\ programs, and a taxonomy for bugs}

% What is a program?
The process of quantum computing involves applying operations on qubits.
Quantum computer scientists use diagrams such as Figure~\ref{fig:bell} to represent sequences of quantum operations.
Looking at Figure~\ref{fig:bell} one sees that quantum programs consist of three conceptual parts:
\begin{enumerate}
\item \textbf{Inputs} to quantum algorithms include quantum initial values for qubits and classical input parameters such as coefficients for rotations.
Getting these inputs to be correct is the focus of Section~\ref{sec:preconditions}.
\item \textbf{Operations} include the specification of how to create an entangled state shown in Figure~\ref{fig:bell}.
Getting these basic operations to be correct is the focus of Section~\ref{sec:basic}.
% Just like in classical computing, a handful of operations are enough to compose all QC programs.
% These universal operations include ``rotations'' on single qubits which alter the probabilities of a qubit's superposition.
% on single qubits, represent rotations on the Bloch sphere
% Pauli matrices: x, y, z
% quaternion pauli matrices
% In the set of universal operations there is also the CNOT operation, which induces entanglement between two qubits.
% Clifford set:
% universal set:
% Hadamard (effect is to create superposition),  CNOT (effect is to create entanglement), S
% Toffoli (CCNOT, reversible AND), and T
Additionally, operations can be further composed according to patterns such as iteration, recursion, and mirroring.
The correctness of these code patterns is the focus of Sections~\ref{sec:iterate},~\ref{sec:recursion}, and~\ref{sec:mirroring}.
\item \textbf{Outputs} of quantum algorithms are the final classical measurement values of qubits such as $m_0$ and $m_1$.
Furthermore, any temporary variables used in the course of a program have to be safely disentangled from the rest of the quantum state and discarded.
Getting these final results to be correct is the focus of Section~\ref{sec:postconditions}.
\end{enumerate}

% Bug taxonomy
%   Classical inputs
% 	Quantum initial values
% 	Basic gates
% 	Composition by iteration
% 	Composition by recursion
% 	Composition by mirroring
% 	Qubit deallocation

% \subsection{Circuits}
% The circuit model
% Circuit depth: number of gates
% Circuit width number of quits
% Quantum volume of 128 bits * 1024 gates would be useful

% Assembly languages for codifying these sequences include OpenQASM~\cite{open_qasm}, Quil~\cite{quil}, and cQASM~\cite{cqasm}, which serve as standard inputs to QC hardware and various simulators~\cite{qx, quest}.
% Qx simulator
% QX Sim gates
% RX, RY, RZ, CR, CZ, CX, S/Phase gates...

% Much research has been devoted to creating high-level QC languages with the goal of providing various useful abstractions for quantum programming~\cite{software_methodology, nature}.
% These high-level languages include imperative languages such as such as Scaffold~\cite{scaffcc}, ProjectQ~\cite{Steiger2018projectqopensource}, Q\#~\cite{q_sharp}, and pyQuil~\cite{quil},
% along with functional languages such as Quipper~\cite{quipper_cacm,quipper}, LIQUi|>~\cite{design_automation}, and QWIRE~\cite{qwire}, just to name a few.
% These languages provide different subsets of useful features, such as the ability to make quantum subroutines, unit testing, polymorphic implementations, quantum-specific syntax, data types, assertions, and efficient simulation.

% What is a bug?
% how can a program can be wrong
% 	Programmer error
% 	the process of getting from theoretical circuit diagrams to code
% 	can be unintuitive
Bugs in quantum programs can crop up in any of these three parts of a QC program
due to mistakes in converting algorithm specifications to program code.
We will give examples of bugs in each of these places using detailed case studies of real quantum programs.
To our knowledge, our work is the first study of such QC program patterns and anti-patterns~\cite{huang_et_al:OASIcs:2019:10196}.
\begin{figure*}[t]
\centering
\includegraphics[width=\linewidth]{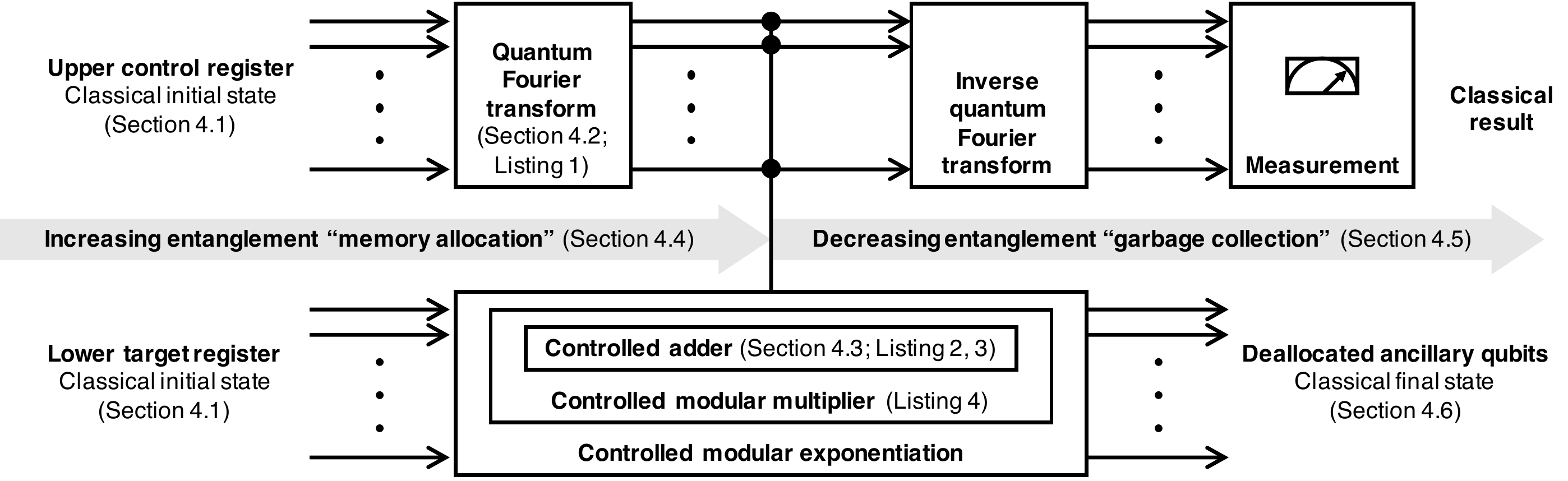}
\Description{A schematic showing what assertions to check for in the different stages of Shor's algorithm}
\caption{
Roadmap for implementing and debugging Shor's algorithm.
}
\label{fig:shors}
\end{figure*}

\section{Our approach to statistical quantum program assertions}
\label{sec:methodology}

Even though it is hard to have as much visibility into quantum program state as is the case in classical computing,
limited but useful assertions checking is possible,
particularly for the purposes of writing correct quantum programs.
In this paper,
we propose using statistical tests on measured outputs as a way to gain visibility into a quantum program (Section~\ref{sec:statistical_assertions}).
We implement several quantum programs in a quantum programming language (Section~\ref{sec:benchmarks}).
Using our ``quantum breakpoints,'' programmers are able to check for expected values at various points in simulated quantum program runs,
allowing them to debug the programs with the aid of our statistical assertions (Section~\ref{sec:breakpoints}).

\subsection{Quantum assertions using statistical tests}
\label{sec:statistical_assertions}

% Debate about what belongs in 03_assertions and what belongs in 06_preconditions and 11_preconditions
% What is common between chi square and chi2 contingency belongs in 03_assertions
% What is not common gets split up

In this subsection,
we preview three basic types of assertions useful in quantum debugging,
and we discuss the mechanics of using statistical tests as quantum assertions.

Following our overview of quantum states, superpositions, and entanglement in Section~\ref{sec:statics},
one already sees that there are three kinds of possible assertions in a quantum program:
\begin{enumerate}
    \item \textbf{Classical assertions:} a quantum variable should take on a deterministic (classical) integer value upon measurement;
    \item \textbf{Superposition assertions:} a quantum variable in superposition should take on a probabilistic distribution of multiple values upon measurement;
    \item \textbf{Entanglement assertions:} two or more quantum variables in an entangled state should take on associated (correlated)\footnote{Correlation is dependence between variables whose magnitude does matter.
That compares with association which is dependence between nominal variables that are merely categories,
and whose magnitude don't matter.} values once they are measured.
\end{enumerate}

Statistical tests use ensembles of multiple measurements to decide to reject hypotheses.
These serve as quantum programming assertions.
With enough measurements, a statistical test is able to tell that an assertion does not hold, indicating a bug in the program.
Otherwise, if the assertion holds,
programmers can proceed cautiously knowing the quantum state so far is consistent with ``no bug,''
given the number of measurements provided to the statistical test.
While this approach is not powerful enough to decisively conclude that a quantum program state is correct,
the debugging experience we share in this paper shows that detecting incorrect states is still useful enough to catch program bugs.
% Our tool offers useful information especially given the constraints on programmers' ability to observe program states in quantum computing.

Specifically, our tool uses the \textbf{chi-square test} to check for classical and superposition quantum states,
and it uses \textbf{contingency table analysis} coupled with the chi-square test to check for entangled states~\cite{numerical_recipes}.
The assertions on classical and superposition quantum states are useful for quantum algorithm precondition checks and for unit testing, discussed in Sections~\ref{sec:preconditions},~\ref{sec:basic},~\ref{sec:iterate}, and~\ref{sec:postconditions}.
Similarly, the assertions on entangled states are useful for checking interaction between qubits,
discussed in Sections~\ref{sec:recursion} and~\ref{sec:mirroring}.

% maybe need to switch to probability not amplitude (done)
\begin{table*}[h]
\centering
\caption{
Correct and incorrect code for rotation decomposition.
Using the Scaffold language~\cite{scaffcc} as an example,
we code out Figure~\ref{fig:decomposition}'s controlled operation U, where U is a rotation in just one axis.
Because only one axis is needed, we can drop either operation A or C, paying attention to the sign on the angles.
Reordering the lines of code or signs results in a rotation in the wrong direction.
}
\begin{tabularx}{\textwidth}{XX|X}
\toprule
\textbf{Correct, operation A unneeded}
&
\textbf{Correct, operation C unneeded}
&
\textbf{Incorrect, angles flipped}\\
\midrule
\texttt{Rz(q1,+angle/2); // C} &
\texttt{CNOT(q0,q1);} &
\texttt{Rz(q1,-angle/2);} \\
\texttt{CNOT(q0,q1);} &
\texttt{Rz(q1,-angle/2); // B} &
\texttt{CNOT(q0,q1);} \\
\texttt{Rz(q1,-angle/2); // B} &
\texttt{CNOT(q0,q1);} &
\texttt{Rz(q1,+angle/2);} \\
\texttt{CNOT(q0,q1);} &
\texttt{Rz(q1,+angle/2); // A} &
\texttt{CNOT(q0,q1);} \\
\texttt{Rz(q0,+angle/2); // D} &
\texttt{Rz(q0,+angle/2); // D} &
\texttt{Rz(q0,+angle/2); // D} \\
\bottomrule
\end{tabularx}
\label{tab:bug_example}
\end{table*}

\subsection{Benchmark QC algorithms for debugging}
\label{sec:benchmarks}

To demonstrate using our assertions framework to debug quantum programs,
we focus this paper on debugging three programs in factoring integers, database search, and quantum chemistry,
each representing a different class of quantum algorithms.

Using the \emph{Shor's integer factoring} quantum algorithm for factoring integers as a centerpiece example throughout Section~\ref{sec:shors},
we show how the structure of quantum programs guides programmers where to put useful quantum assertions.
We propose a complete taxonomy of where bugs can take place,
and show assertions can catch all of the categories of bugs.

Then, using the \emph{Grover's database search} and a \emph{quantum chemistry} problem as additional case studies in Section~\ref{sec:algorithms},
we discuss how different classes of quantum algorithms present different opportunities and challenges for debugging.

\subsection{Simulation and assertions checking methodology}
\label{sec:breakpoints}

% What we built
We implement the programs in the Scaffold quantum programming language developed by our research group~\cite{scaffcc},\footnote{\url{https://github.com/epiqc/ScaffCC}}
now augmented with the ability to specify and check for assertions.
The assertions instruct the compiler where to stop execution and measure qubit states.
Since premature measurement destroys the quantum state,
the assertions effectively terminate the quantum program,
splitting the quantum program into multiple breakpoints.

% Scaffold does not like to run in parallel
% copy to mkdir -p /scratch/$USER ; cd /scratch/$USER
% copy back to SLURM SUBMIT DIR
Our tool uses the ScaffCC compiler to compile Scaffold code with assertions into multiple versions of OpenQASM,
a QC assembly language~\cite{open_qasm}.
Each version of the compiled program has the program execution up to the quantum breakpoint,
followed by an early measurement and assertions on expected values for the quantum variables.

Then,
our tool simulates an ensemble of executions for each of the programs ending at each breakpoint,
using the QX quantum simulator~\cite{qx} running on a cluster compute infrastructure.

Finally, the tool gathers the results from the simulations to check for the assertions at each breakpoint.
The measurement results feed into statistical tests,
which check if the quantum variables have values that are consistent with 
being in one of classical, superposition, or entangled states.
If the statistical tests reject the null hypotheses,
that indicates the assertions were violated,
which means there is a bug in the quantum program or that the assertion was an incorrect constraint.

Given the rapid growth of QC infrastructure,
programmers now have the chance to test a variety of quantum algorithms written in many languages~\cite{comparison}.
To validate our overall approach,
we cross-validated our quantum programs and simulation results against equivalent programs written in other quantum programming languages,
such as LIQUi|>~\cite{design_automation}, ProjectQ~\cite{emulation, Steiger2018projectqopensource}, and Q\#~\cite{q_sharp}.

From this detailed debugging experience spanning multiple algorithms, languages, and simulators,
we are able to concretely describe for the first time what types of bugs may crop up in quantum programs,
and how assertions can aid in the debugging of quantum programs.
As an added contribution, Section~\ref{sec:algorithms} discusses how language features of different QC programming languages can aid with the placement of quantum assertions,
or otherwise prevent bugs in the first place.

% cRz / QFT are minimal examples here
\begin{figure}[t]
\centering
\includegraphics[width=\columnwidth]{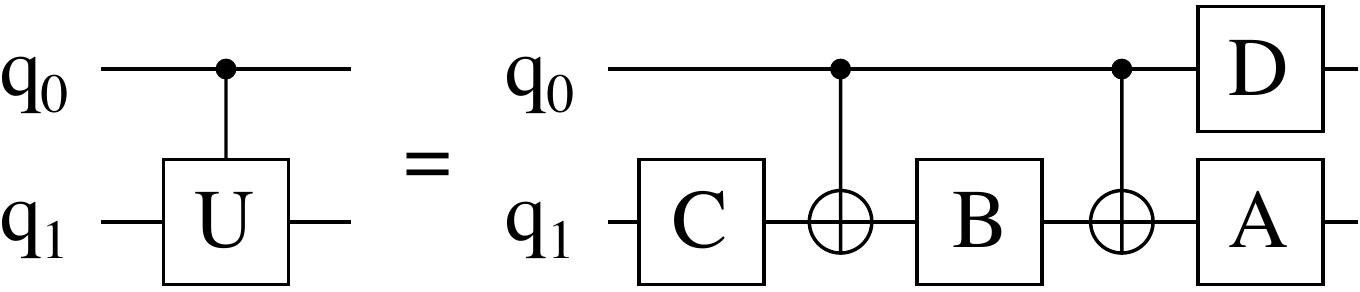}
\Description{Schematic for decomposing a controlled-U operation two qubits, into more elementary single-qubit and double-qubit gates.}
\caption{
Decomposition of a simple QC program.
Time flows left to right, showing sequences of operations applied to qubits $q_0$ and $q_1$.
The left symbol is a \emph{controlled} arbitrary operation $U$.
Whether the operation $U$ applies to the \emph{target} qubit $q_1$ is dependent on the value of the \emph{control} qubit $q_0$.
The diagram on the right shows the decomposition into the equivalent sequence of more basic operations.
The basic operations include single-qubit \emph{rotations} $A$ through $D$ that alter the probability distribution of qubit values.
The operations also include two two-qubit \emph{controlled-NOT} operations that flip the target qubit (denoted $\oplus$) contingent on the value of the control qubit (denoted $\bullet$)~\cite{nielsen_chuang}.}
\label{fig:decomposition}
\end{figure}

\section[QC debugging and assertions: Shor's algorithm case study]{QC debugging and assertions:\\ Shor's algorithm case study}
\label{sec:shors}

Using the Shor's quantum algorithm for factoring integers as a concrete case study,
we show how the structure of the quantum program shown in Figure~\ref{fig:shors} aids programmers in using assertions to debug quantum programs.
To defend against bugs in quantum program input, operations, and outputs,
programmers can write assertions that check for preconditions, invariants, and postconditions.
These constraints aid in the process of bringing up the program from unit tests to overall integration tests.

% How the workshop paper observations contribute to this being a self-contained, synergistic framework

% top down programmer modules, reversible computing, controlled operation language level annotation
%     drives placement of assertions
% data flow analysis of entanglement
% entanglement anomaly detection
% ability to tell the effect of CNOTs and quality of CNOTs with noise
% detection of hardware defects for two qubit gates

Shor's factorization algorithm uses a quantum computer to factor a composite number in polynomial time complexity, providing exponential speedup relative to the best known classical algorithms~\cite{shor}.
% Shor algorithm: exponential speedup relative to current best algorithm
% Quantum Shor algorithm from 1994 can do this has cost (log N)^2 with high probability
% Factoring a huge composite number N, find (prime) factors of N; best case now is beyond polynomial
% Note that the best case complexity of a classical (non quantum) algorithm is not known; in future there may be polynomial algorithm
The algorithm works by estimating eigenvalues of a matrix,
where the matrix is generated from the exponentiation of an integer representing a trial factor.
The arithmetic is done in modular space with the modulus $N$ set to be the integer one wants to factor.
Here, we replicate results for factoring $N=15$, the simplest example~\cite{experimental_demo_shor}~\cite[p. 235]{nielsen_chuang},
by following an example for an implementation that minimizes the qubit cost~\cite{beauregard}.
% Shor algorithm specific implementations: BCDP, F, D
Once the quantum part of Shor's algorithm is done finding the eigenvalues,
those values are useful in a classical post-processing algorithm to find 3 and 5, the factors of 15.

We focus on debugging the Shor's factoring algorithm because it features in a single overall algorithm several important primitives (kernels) and program patterns common to many quantum algorithms.
The primitives invoked in Shor's algorithm include order finding, eigenvalue estimation, state preparation, phase estimation, and quantum Fourier transform.
Our assertions and debugging techniques apply to several other QC applications that invoke similar primitives and patterns.
\subsection{Classical and superposition precondition assertions on quantum initial values}
\label{sec:preconditions}

Correct implementation and execution of QC programs begins with the right input states.
Given their importance, it is worthwhile to check these preconditions by running or simulating programs up to the entry point of subroutines,
and performing a premature measurement to check for these anticipated states.

% which includes HHL for linear algebra

\paragraph{\emph{\textbf{Bug type 1: Incorrect quantum initial values.}}}
The type of quantum initial state that an algorithm needs depends on the type of algorithm.
For eigenvalue estimation algorithms such as Shor's algorithm,
the two major pieces of quantum data are an upper register and a lower register (far left of Figure~\ref{fig:shors}):
the upper register participates in a phase estimation subroutine;
while the lower register is scratch space to implement a mathematical function such as modular exponentiation.

% The lower register is a temporary variable, which are called ancillary qubits in quantum computation.
Here, the \emph{lower target register} needs to be initialized to a strictly \emph{classical integer} value, such as `1'.
That means that if a quantum computer measured the qubit encoding the least significant bit of the quantum variable, it should return `1',
while the measurements on the other qubits of the variable should return `0'.

On the other hand, the \emph{upper control register} needs to be initialized to a uniform  \emph{superposition} of values.
More concretely, if the upper register consists of, for example, 3 qubits, the measurement of the upper register at the beginning of the algorithm should return the eight values `000', `001', ... `111' with equal probability.
That uniform superposition of values is created by the quantum Fourier transform (QFT).
The QFT operation has the effect of taking integer inputs,
and re-encoding them as quantum values that are distinct from each other by a quantum property known as \emph{phase}.

The classical value in the upper register and uniform superposition in the lower register are the preconditions for Shor's algorithm.
Other types of preconditions are possible for other types of algorithms.
For example, quantum communications protocols often need entangled states as initial conditions.

\paragraph{\emph{\textbf{Defense type 1: Assertion checks for classical and superposition preconditions.}}}
Our tool checks for both classical and superposition states using chi-square statistical tests on measured values.

To test for \emph{classical integer} values,
our tool gives the chi-square test the hypothesis that the distribution is unimodal with a peak at the expected value.
% (difficult when many counts of bins are small)
If the test returns a small $p$-value ($\leq0.05$),
then the null hypothesis is rejected,
meaning the initial state cannot be the expected value,
indicating a violation of the precondition.
If, on the other hand, the test returns a large $p$-value, typically close to 1.0,
then the input state is consistent with being the expected value,
though programmers cannot completely rule out a precondition violation.
If there actually was a bug, programmers would only be able to detect the precondition violation using more measurements.
% it has to be pretty bad before null hypothesis is rejected and you can conclude input is bad

To test for \emph{superposition} quantum states of $n$-qubits,
the chi-square test uses as its hypothesis that the measurements should be a uniform distribution across all $2^n$ integer values.
If the superposition precondition is violated, and there are sufficient measurements,
the values would be concentrated enough for the chi-square test to reject the null hypothesis and raise an exception.
% needs to have large p value for uniform distribution
% Chi needs to be small for uniform distribution

% Scalability
% How many samples do you need to catch a bug
% Qubit count X error rate X $p$-value

In prior work,
the Q\# quantum programming language has support for assertion checks for \emph{integer} values,
and is able to check for such assertions in simulations of quantum programs~\cite{q_sharp}.
% Q\# assert and assertprob
% assert prob only supports integer assertions
% https://docs.microsoft.com/en-us/quantum/libraries/standard/testing?view=qsharp-preview
To our knowledge, this paper is the first proposal for quantum assertions on \emph{superposition} and (later in this paper) \emph{entangled} quantum states.
Furthermore, this is the first work to discuss using statistical tests to check for these hypotheses.
\lstdefinestyle{scaffold}{
    frame=single,
    language=C,
    morekeywords={qbit,PrepZ,H,assert_classical,assert_superposition,assert_entangled,assert_product},
    backgroundcolor=\color{white},
    commentstyle=\color{teal},
    keywordstyle=\color{blue},
    numberstyle=\footnotesize\color{darkgray},
    stringstyle=\color{orange},
    basicstyle=\ttfamily,
    breakatwhitespace=false,
    breaklines=true,
    captionpos=b,
    keepspaces=true,
    numbers=left,
    numbersep=5pt,
    showspaces=false,
    showstringspaces=false,
    showtabs=false,
    tabsize=2,
}

\begin{figure}[t]
\begin{lstlisting}[
style=scaffold,
caption={Test harness for quantum Fourier transform.},
label=lst:qft_test
]
#include "QFT.scaffold"
#define width 4 // number of qubits
int main () {

  // initialize quantum variable to 5
  qbit reg[width];
  for ( int i=0; i<width; i++ ) {
    PrepZ ( reg[i], (i+1)%2 ); // 0b0101
  }

  // precondition for QFT:
  assert_classical ( reg, width, 5 );

  QFT ( width, reg );

  // postcondition for QFT &
  // precondition for iQFT:
  assert_superposition ( reg, width );

  iQFT ( width, reg );

  // postcondition for iQFT:
  assert_classical ( reg, width, 5 );
}
\end{lstlisting}
\end{figure}

\begin{figure*}[t]
\begin{lstlisting}[
style=scaffold,
caption={Controlled adder subroutine using QFT.},
label=lst:controlled_adder
]
// outputs b <= a+b, where a is a `width' bit constant integer
// b is an integer encoded on `width' qubits in Fourier space
module cADD (
  const unsigned int c_width, // number of control qubits
  qbit ctrl0, qbit ctrl1, // control qubits
  const unsigned int width, const unsigned int a, qbit b[]
) {
  for ( int b_indx=width-1; b_indx>=0; b_indx-- ) {
    for ( int a_indx=b_indx; a_indx>=0; a_indx-- ) {
      if ( (a>>a_indx) & 1 ) { // shift out bits in constant a
        double angle = M_PI / pow ( 2, b_indx-a_indx ); // rotation angle
        switch (c_width) {
          case 0: Rz ( b[b_indx], angle ); break;
          case 1: cRz ( ctrl0, b[b_indx], angle ); break;
          case 2: ccRz ( ctrl0, ctrl1, b[b_indx], angle ); break;
}}}}}
\end{lstlisting}
\end{figure*}

\begin{figure}[t]
\begin{lstlisting}[
style=scaffold,
caption={Test harness for controlled adder subroutine.},
label=lst:controlled_adder_test
]
#include "cADD.scaffold" // see Listing 2
#define width 5 // number of qubits
int main () {
  // control qubits unimportant here
  qbit ctrl[2];
  PrepZ ( ctrl[0], 0 );
  PrepZ ( ctrl[1], 0 );

  // initialize quantum variable to 12
  const unsigned int b_val = 12;
  qbit b[width];
  for ( int i=0; i<width; i++ ) {
    PrepZ ( b[i], (b_val>>i)&1 );
  }
  assert_classical( b, width, 12 );

  // perform the addition
  QFT ( width, b );
  const unsigned int a = 13;
  cADD (0,ctrl[0],ctrl[1],width,a,b);
  iQFT ( width, b );

  // assert a+b = 12+13 = 25
  assert_classical ( b, width, 25 );
}
\end{lstlisting}
\end{figure}

\subsection{Unit tests for a library of subroutines}

Now that we have made sure the quantum initial states are valid,
the next step in programming the Shor's algorithm is to build up the algorithm operations.
We do so starting from elementary operations,
which we exhaustively validate against their closed form solutions,
and against implementations in other languages.
Then we compose the elementary operations following code patterns---iterations, recursion, and mirroring---and test the composite subroutines using assertions (Sections~\ref{sec:iteration},~\ref{sec:recursion},~\ref{sec:mirroring}).

\paragraph{\emph{\textbf{Bug type 2: Incorrect operations and transformations.}}}
\label{sec:basic}
In order to correctly implement Shor's algorithm,
programmers first have to build up the quantum subroutines such as the \emph{controlled rotation} subroutine depicted in  Figure~\ref{fig:decomposition}.
This subroutine is the building block for QFT and adder routines in Shor's algorithm (modules in Figure~\ref{fig:shors}).
Typically this task consists of translating quantum circuit diagrams, such as Figure~\ref{fig:decomposition},
into quantum program code.
Sometimes, programmers do not even have quantum circuit diagrams and must instead start with equation descriptions for the operations they need.
This process of converting specifications to program code is unintuitive and tricky.
For example,
Table~\ref{tab:bug_example} lists multiple ways to code the decomposition of the controlled rotation,
including a buggy one where a small mistake leads to the wrong operation.

% Shors modadd submodule works even when underlying cRz is wrong, without phase correction

% Code readability vs optimized?
% Just the Grover's: compare gate count for canonical vs. tricky optimized
% Nielsen Chuang Pg 184 figure 4.10. innermost control U
% or p258 central most z between "toffolis"

\paragraph{\emph{\textbf{Defense type 2: Assertion checks for unit testing.}}}
\label{sec:modularization}
An obvious defense against coding mistakes in basic subroutines (such as controlled rotation, QFT, and addition subroutines) is to use a library of shared code.
Doing so helps ensure program correctness by allowing programmers to exhaustively validate small subroutines,
in order to bootstrap larger subroutines.
Unit testing is especially important in QC as running or simulating large quantum programs is costly, making larger scale integration tests impossible.

% QFT example
As a concrete example, we use precondition and postcondition assertion checks inside a test harness to validate the QFT subroutine, another important building block.
As shown in Listing~\ref{lst:qft_test},
first the program prepares a classical integer state (Lines 5-9).
Then, the program checks as a precondition of the QFT subroutine that the input is a classical integer value,
in this case `5' (Line 12).
The corresponding postcondition of the QFT subroutine is that the output should be a uniform superposition if the program collapses the quantum state and measure the values at that point (Line 18).

While these simple constraints are not enough on their own to validate that the QFT implementation and its sub-components are correct,
they are valuable lightweight sanity checks.
For the QFT subroutine, additional validation comes from cross checking its outputs against closed form mathematical solutions,
and against implementations in other languages.

\subsection[Numeric assertion checks for composing gates with iterations]{Numeric assertion checks for\\ composing gates with iterations}
\label{sec:iteration}

From the basic subroutines,
programmers typically compose the subroutines into quantum programs using patterns including iterations, recursion, and mirroring.
Here we focus on iterations,
a pattern commonly invoked in code related to the QFT for the purpose of manipulating qubits that represent numbers.
Our tool can catch bugs in iteration code using assertions on integer inputs to and outputs from subroutines.

\paragraph{\emph{\textbf{Bug type 3: Incorrect composition of operations using iteration.}}}
\label{sec:iterate}
Now that we have validated code for the controlled rotation and QFT subroutines,
the next more complex subroutine is the controlled adder,
which is itself a subroutine for the modular exponentiation part of Shor's algorithm (bottom module in Figure~\ref{fig:shors}).
Listing~\ref{lst:controlled_adder} shows the iteration code for the constant-value adder,
showing tricky places in Lines 8 through 11 where bugs can crop up.
These possible bugs include indexing errors in the two-dimensional loop,
bit shifting errors, endian confusion, and mistakes in rotation angles.\footnote{One of the trickiest aspects of quantum programming is properly keeping track of how quantum variables map to qubit assignments.
One way to prevent bugs altogether in this kind of code is to introduce QC data types for numbers,
providing greater abstraction than working with raw qubits.
% ProjectQ offers quint, qufixed, qufloat, interpretations of query
For example, ProjectQ has quantum integer data types~\cite{Steiger2018projectqopensource},
while Q\#~\cite{q_sharp} and Quipper~\cite{quipper, quipper_cacm} offer both big endian and little endian versions of subroutines involving iterations.
These QC data types permit useful operators (e.g., checking for equality) that help with debugging and writing assertions.}

\paragraph{\emph{\textbf{Defense type 3: Assertion checks for classical intermediate states.}}}
\label{sec:numerical}
Our tool's assertions on classical integer values allows for unit testing of code that involve iterations.
As an example in Listing~\ref{lst:controlled_adder_test},
programmers can write assertions on the inputs (Line 15) and outputs (Line 24) of the controlled adder subroutine.
With these assertions programmers can catch coding mistakes made in its constituent subroutines.
For example the bug involving the incorrect version of the rotation operation in Table~\ref{tab:bug_example} is caught here when the output assertion returns $p\mathrm{-value}=0.0$,
indicating the addition did not work as expected,
due to a bug inside the controlled adder.
\begin{figure}[t]
\begin{lstlisting}[
style=scaffold,
caption={Test harness for the controlled modular multiplier subroutine.},
label=lst:controlled_modular_multiplier_test
]
#include "cMODMUL.scaffold"
#define width 5 // number of qubits
#define N 15 // number to factor

// CALCULATE: b <= a*x+b mod N
int main () {

  // control qubit in superposition
  qbit ctrl[1];
  PrepZ ( ctrl[0], 1 );
  H ( ctrl[0] );

  // initialize x variable to 6
  const unsigned int x_val = 6;
  qbit x[width];
  for ( int i=0; i<width; i++ ) {
    PrepZ ( x[i], (x_val>>i)&1 );
  }
  assert_classical ( x, width, 6 );

  // initialize b variable to 7
  const unsigned int b_val = 7;
  qbit b[width];
  for ( int i=0; i<width; i++ ) {
    PrepZ ( b[i], (b_val>>i)&1 );
  }
  assert_classical ( b, width, 7 );

  // ancillary qubits unimportant here
  qbit ancilla[1];
  PrepZ ( ancilla[0], 0 );

  // perform modular multiplication
  const unsigned int a = 7;
  cMODMUL ( ctrl[0], width, a, x, b, N, ancilla[0] );

  assert_entangled( ctrl,1, b,width );

  // inverse modular multiplication
  const unsigned int a_inv = 13;
  cMODMUL ( ctrl[0], width, a_inv, x, b, N, ancilla[0] );

  assert_product( ctrl,1, b,width );
}
\end{lstlisting}
\end{figure}

\begin{figure}[t]
\centering
\includegraphics[width=0.5\columnwidth]{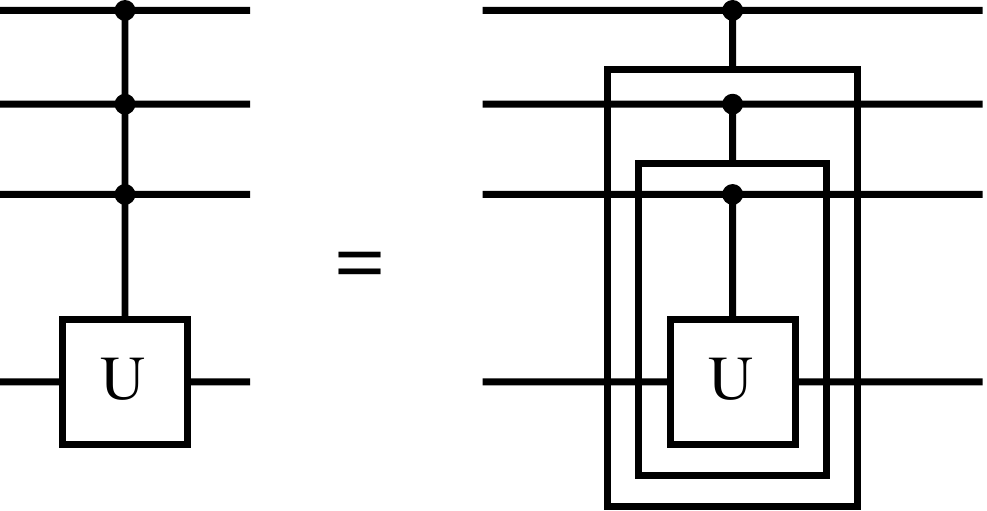}
\Description{A schematic showing that multiply-controlled operations have a nested structure.}
\caption{Controlled operations with multiple control qubits result in recursive code patterns.}
\label{fig:recursion}
\end{figure}

\subsection[Entanglement assertion checks for composing gates with recursion]{Entanglement assertion checks for\\ composing gates with recursion}
\label{sec:recursion}

The next two types of bugs in quantum programs have to do with two more ways to compose basic operations, both of which have to do with the interaction between quantum variables;
i.e., between two or more sets of qubits.
That is in contrast to the previous two types of bugs in basic operations and iterating operations, which generally act on single variables (where the variables may comprise multiple qubits).

In quantum computing, the interaction between variables takes place through entanglement.
For example, in Figure~\ref{fig:shors}, the upper and lower registers interact when they are entangled through the controlled modular exponentiation operation.
If two variables are entangled when a quantum computer measures them,
the classical values that they collapse to will be correlated.
Using statistical tests on the measurement results,
programmers can write assertions to check whether variables are entangled as expected.

\paragraph{\emph{\textbf{Bug type 4: Incorrect composition of operations using recursion.}}}
Entanglement is achieved using controlled operations,
which is a common pattern in quantum programs that involves performing operations (e.g., modular multiply),
contingent on a set of qubits known as control qubits.
These controlled operations correspond to using recursion to compose basic operations.
A multiply-controlled rotation, for example, is just a controlled rotation that is itself controlled by other qubits (Figure~\ref{fig:recursion}).

The process of coding recursive operation patterns may introduce bugs.
That is because quantum algorithms often need varying numbers of control qubits in different parts of the algorithm,
leading to replicated code from multiple versions of the same subroutine differing only by the number of control qubits.
An example appears in Listing~\ref{lst:controlled_adder},
where the addition operation is contingent on control qubits taken as parameters in Lines 4 and 5.
Depending on how many control qubits are needed,
the switch statement in Lines 12 through 15 applies the correct operation.
The specific bug we are going to demonstrate catching next is if a programmer made a mistake in Line 15,
where they accidentally use \texttt{ctrl1} twice instead of \texttt{ctrl0},
causing a mistake in how the control qubits are routed.

\paragraph{\emph{\textbf{Defense type 4: Assertion checks for entangled intermediate states.}}}
\label{sec:entanglement}
Programmers can check for these types of bugs in recursive code patterns for controlled operations using entanglement assertions,
a new kind of quantum assertion that we introduce to test for dependence between measured values.

% Hadamard controlled-X leading to Bell state mini tutorial
As a very simple example,
we show how the entanglement assertion check works on the simplest example of entangled states.
In the Bell state creation circuit we showed in Figure~\ref{fig:bell},
% the control qubit is first put into superposition using a Hadamard gate.
% The control qubit then is the condition for a CNOT gate,
% which negates the value of the target qubit,
% contingent on the value of the control qubit.
the state of the two qubits $Q$ in location (D) of that diagram are in a \emph{Bell state},
a minimal example of an entangled state between qubits.
The measurement results $m_0$ and $m_1$ are maximally correlated---either both return `0' or both return `1'.
Using such observations one can build a contingency table:

\centerline{
\begin{tabular}{ c r|p{1.25cm}p{1.25cm} } 
\toprule
\multicolumn{2}{c|}{\multirow{2}{*}{\textbf{Probability}}} & \multicolumn{2}{c}{$m_0$ measurement} \\
&& 0 & 1 \\
\hline
\multirow{1}{*}{$m_1$} & 0 & 1/2 & 0 \\
\multirow{1}{*}{measurement} & 1 & 0 & 1/2 \\
\bottomrule
\end{tabular}
}

% We can use contingency table analysis,
% also known as cross-tabulation analysis,
% to determine whether two qubits are entangled,
% and use this simple statistical test for association as a kind of assertion check.
% To do so, we would build a contingency table of the measurement outcomes of the control and target qubits.

Next we again use a chi-square statistical test on the table to determine a contingency coefficient.
If the $p$-value is small ($\leq0.05$),
as is the case for this table,
then the test rejects the null hypothesis and concludes the observations must be correlated,
and therefore the quantum variables were entangled when they were measured.
On the other hand,
if the $p$-value fails to be significant,
then the observations are consistent with the variables being independent and unentangled.
% phi coefficient

% Note that the entanglement assertion is a bit distinct from the previous two assertions for classical integer and superposition states.
% In order to check for entangled states we are actually asserting that variables are not independent.

These entanglement assertions are powerful tools for catching bugs such as our example bug of mistaken control qubits in the controlled adder subroutine.
Our tool catches the bug using entanglement assertions in the controlled modular multiplier test harness,
shown in Listing~\ref{lst:controlled_modular_multiplier_test}.
To prepare the contingency table,
the programmer only needs to identify pairs of quantum variables that should be entangled with each other using the \texttt{assert\_entangled} statement (Line 37),
which takes four parameters specifying the control and target quantum variables and their bitwidths.
Then, our debugging tool keeps track of which qubits those specified variables correspond to.
The simulator then does an early measurement of the qubits for both variables.
The debugging tool then maps the measurement results into columns and rows of a contingency table automatically, and a chi-square test checks to make sure the control qubits have an effect on whether the multiplier acts on the target qubits.

If the controlled add operation is bug-free, with the control qubits correctly routed,
the first assertion returns $p\mathrm{-value}=0.0005$ for an ensemble size of 16,
indicating the control and target register values are entangled at the point of the assertion.
That means that whichever way the control qubit collapses out of its superposition state,
it correctly controls whether the multiplication works on the target register.
On the other hand, if the control qubits are routed incorrectly,
the first assertion returns $p\mathrm{-value}=0.121$ for an ensemble size of 16.
This indicates the control register value is not correctly toggling the operation of the multiplier,
hinting the bug must be somewhere inside the multiplier implementation.
\begin{table}[t]
\centering
\caption{Correct classical input $a$ and $a^{-1}$ to Shor's algorithm for factoring 15, using 7 as a guess.}
\begin{tabularx}{\linewidth}{ r|XXXXX } 
\toprule
\textbf{$k$, the algorithm iteration} & \textbf{0} & \textbf{1} & \textbf{2} & \textbf{3} & \textbf{\ldots} \\
\midrule
$a=7^{2^k} \mod 15$ & 7 & 4 & 1 & 1 & \ldots \\
$a^{-1}$; $a\times a^{-1} \equiv 1 \mod 15$ & 13 & 4 & 1 & 1 & \ldots \\
\bottomrule
\end{tabularx}
\label{tab:shor_inputs}
\end{table}

\begin{table*}
\centering
\caption{Probability of measuring values of outputs and ancillary qubits of Shor's algorithm, with incorrect inputs ($a^{-1}=12$ instead of $13$ on first iteration).
If the ancillary qubits collapse to zero on measurement, the algorithm still succeeds, returning correct outputs of $0$, $2$, $4$, $6$~\cite[p. 235]{nielsen_chuang}.
However, the possibility of measuring non-zero for the ancillary qubits indicates a bug.
% So we would assert ancillary qubits = 0 as a postcondition.
}
\begin{tabularx}{\linewidth}{ c r|XXXXXXXX } 
\toprule
\multicolumn{2}{c|}{\multirow{2}{*}{\textbf{Probability}}} & \multicolumn{8}{c}{\textbf{Output measurement}} \\
&& 0 & 1 & 2 & 3 & 4 & 5 & 6 & 7 \\
\midrule
\multirow{3}{*}{\textbf{Ancillary}} & 0 & 1/8 & 0 & 1/8 & 0 & 1/8 & 0 & 1/8 & 0 \\
\multirow{3}{*}{\textbf{qubits}} & 2 & 1/64 & 1/64 & 1/64 & 1/64 & 1/64 & 1/64 & 1/64 & 1/64 \\
\multirow{3}{*}{\textbf{measurement}} & 7 & 1/64 & 1/64 & 1/64 & 1/64 & 1/64 & 1/64 & 1/64 & 1/64 \\
& 8 & 1/64 & 1/64 & 1/64 & 1/64 & 1/64 & 1/64 & 1/64 & 1/64 \\
& 13 & 1/64 & 1/64 & 1/64 & 1/64 & 1/64 & 1/64 & 1/64 & 1/64 \\
\bottomrule
\end{tabularx}
\label{tab:shor_outputs}
\end{table*}

\subsection[Product state postcondition assertions for composing gates with mirroring]{Product state postcondition assertions for\\ composing gates with mirroring}
\label{sec:mirroring}

Contingency table analysis is also useful for checking for the third and final kind of pattern in quantum programs,
the correct mirroring of operations.
The reason this pattern appears in quantum programs is to allocate and deallocate qubits within a quantum subroutine,
analogous to the allocation and garbage collection of memory in classical programs.
For example,
the Shor's algorithm in Figure~\ref{fig:shors} can be seen as allocating the bottom register of qubits (known as ancillary qubits) in the left half of the algorithm,
performing the modular exponentiation,
and then deallocating the bottom register of qubits in the right half of the algorithm.
Product state assertions validate that the deallocation of these ancillary qubits is done correctly.

\paragraph{\emph{\textbf{Bug type 5: Incorrect composition of operations using mirroring.}}}
In order to garbage collect ancillary qubits in quantum programs,
programmers need to reverse all the operations they applied to the qubits.

The reason programmers have to do so is because garbage collection is different in quantum computing compared to that in classical computing.
In classical computing,
programmers can simply mark any memory as unneeded in order to free it,
and that memory would be rewritten some time later in program execution.
But in quantum computing,
qubits can be entangled and therefore cannot be treated as independent pieces of information.
Suppose a program is done with using the lower register in Figure~\ref{fig:shors},
but they remain entangled with the upper register qubits.
Then anything that happens to the ancillary qubits,
such as measurement,
re-initialization,
or lapsing into incoherence,
can have unintended effects on the output qubits in the upper register that the program user does care about.

To prevent these unintended side effects,
programmers have to carefully undo any entanglement they have built up between qubits.
To do this, programmers perform inverse operations in backward order from the order they originally performed them.
This process is called \emph{uncomputation}~\cite{Kaye:2007:IQC:1206629,nielsen_chuang,software_methodology}.
After uncomputation, ancillary qubits should be properly untangled from the rest of the program state, and are truly ready for reuse.

% \footnote{
% Reversing operations is possible because all QC operations mathematically entail multiplying quantum states by unitary matrices, which have an inverse.
% An important caveat is the matrices must be ``unitary'', which enforces that the operations preserve the total probability of all the measured outcomes.
% so the amplitude does not change when you chain these
% UUT = UTU = I
% where T is conjugate transpose
% reversibility is important; it’s a key distinction between classical computing and QC; all operations are reversible
% },

This process of uncomputation can be tricky if done manually.
Take for example the controlled adder subroutine shown in Listing~\ref{lst:controlled_adder}.
Uncomputing the addition operation would need an inverse adder counterpart to the controlled adder.
The code for the inverse adder would have each of the iterations in Lines 8 and 9 iterated in reverse order,
and would have the rotation angles used in Lines 13 through 15 negated.
Bugs in these inverse operations would impact the qubit deallocation process.

\paragraph{\emph{\textbf{Defense type 5: Assertion checks for product state postconditions.}}}
As a counterpart to entanglement assertions,
our tool offers product state assertions to make sure that ancillary qubits and output qubits are in a product state,
meaning they have no entanglement.
This kind of assertion would make sure that code for the pattern of mirroring operations is correct.

We demonstrate the use of product state assertions also in Listing~\ref{lst:controlled_modular_multiplier_test}.
Following the controlled modular multiplier in Line 35,
the program reverses that operation in Line 41.
The way the program invokes the inverse operation in Line 41 is by multiplying by the modular inverse.
In the example here, $7\times13\equiv1 \; \mathrm{mod} \; 15$,
so multiplying by $a^{-1}=13$ inverts the operation of multiplying by $a=7$.
With the correct inverse computation, the \texttt{assert\_product} statement in Line 43 returns $p\mathrm{-value}=1.0$,
consistent with no entanglement between the upper control register and the bottom target register,
indicating the bottom register is properly deallocated.

If on the other hand the program mistakenly multiplies by any number that is not the modular inverse,
for example $a^{-1}=12$,
then the assertion returns $p\mathrm{-value}=0.0005$ (for an ensemble size of 16) indicating the two registers are still incorrectly entangled,
meaning the bottom register was not correctly deallocated,
which hints to a bug in the mirrored code.

% so long as the numbers are coprime
% 2 and 8
% 4 and 4
% 7 and 13
% 8 and 2
% 11 and 11
% 13 and 7
% 14 and 14

\subsection[Classical postcondition assertions on deallocated ancillary qubits]{Classical postcondition assertions on\\ deallocated ancillary qubits}

Finally, we are ready to run the Shor's algorithm in an overall integration test.
To run Shor's algorithm, the programmer has to feed the algorithm pairs of modular inverse numbers as its input.
For example, Table~\ref{tab:shor_inputs} shows the input pairs for factoring 15, using 7 as a trial divisor.
Then, the algorithm should return 0, 2, 4, or 6, each with equal probability, from measuring the upper register~\cite[p. 235]{nielsen_chuang}.
These numbers would go into a classical post-processing algorithm to find the factors of $3\times5=15$.

Typically,
programmers would only measure the upper register of qubits (Figure~\ref{fig:shors}) that carry program output,
and ignore the bottom register of qubits as they are merely ancillary qubits and should carry no information.
However, when a programmer is debugging a quantum program,
these ancillary qubits often carry useful side channel information that informs the programmer whether the main outputs are valid.
Our tool checks for this information using classical assertions on the expected values for these deallocated ancillary qubits.

\paragraph{\emph{\textbf{Bug type 6: Incorrect classical input parameters.}}}
\label{sec:deallocation}
The final bug we study for Shor's algorithm stems from giving wrong input parameters to an otherwise correctly written quantum program.
These mistakes can be difficult to debug,
even though the bug is entirely in the classical inputs to the algorithm.

The specific mistake is the programmer supplies wrong pairs of numbers as modular inverses for the algorithm.
Instead of using $(a,a^{-1})=(7,13)$ for the first iteration in Table~\ref{tab:shor_inputs},
the programmers gives a wrong pair of numbers $(7,12)$.
We show our tool can debug this problem using assertions.

\paragraph{\emph{\textbf{Defense type 6: Assertion checks for classical postconditions.}}}
\label{sec:postconditions}
The outputs of Shor's algorithm for this incorrect pair of inputs is recorded in Table~\ref{tab:shor_outputs}.
The table is a contingency table showing the joint probability for the output measurement and the ancillary qubit measurement.
The table shows the ancillary qubits collapse to a non-zero value with probability $1/2$,
which is incorrect because they should always return to their initial value of $0$ after appropriate uncomputation.
This symptom is to be expected because the incorrect pair of modular inverses fed to the algorithm has caused incorrect inversion of the multiplication operation inside the algorithm.

The programmer can use a classical assertion as a postcondition check on the deallocated ancillary qubits.
The program should assert that the ancillary qubits should return their initial value of $0$.
If the postcondition assertion fails,
the programmer knows there was a bug in the deallocation of qubits and therefore the outputs may be wrong.
If the postcondition succeeds, then the Shor's factoring algorithm returns valid outputs.

\newcommand{\specialcell}[2][c]{%
  \begin{tabular}[#1]{@{}l@{}}#2\end{tabular}}

\begin{table*}
\centering
\caption{Grover's amplitude amplification subroutine in two languages, showcasing QC-specific language syntax for reversible computation (rows 2 \& 6) and controlled operations (rows 3 \& 5), exposing structure that can guide placing assertions.}
\begin{tabularx}{\linewidth}{ r|XX }
\toprule
& \textbf{Scaffold (C syntax)~\cite{scaffcc}} & \textbf{ProjectQ (Python syntax)~\cite{Steiger2018projectqopensource}} \\
\midrule

\textbf{1}&

\specialcell{
\texttt{int j;}\\
\texttt{qbit ancilla[n-1]; // scratch register}\\
\texttt{for(j=0; j<n-1; j++) PrepZ(ancilla[j],0);}
}&

\specialcell{
\texttt{\# reflection across}\\
\texttt{\# uniform superposition}
}

\\\hline

\textbf{2}&

\specialcell{
\texttt{// Hadamard on q}\\
\texttt{for(j=0; j<n; j++) H(q[j]);}\\
\texttt{// Phase flip on q = 0...0 so invert q}\\
\texttt{for(j=0; j<n; j++) X(q[j]);}
}&

\specialcell{
\texttt{with Compute(eng):}\\
\texttt{~~~~All(H) | q}\\
\texttt{~~~~All(X) | q}
}

\\\midrule

\textbf{3}&

\specialcell{
\texttt{// Compute x[n-2] = q[0] and ... and q[n-1]}\\
\texttt{CCNOT(q[1], q[0], ancilla[0]);}\\
\texttt{for(j=1; j<n-1; j++)}\\
\texttt{~~~~CCNOT(ancilla[j-1], q[j+1], ancilla[j]);}
}&

\specialcell{
\texttt{with Control(eng, q[0:-1]):}
}

\\\midrule

\textbf{4}&

\specialcell{
\texttt{// Phase flip Z if q=00...0}\\
\texttt{cZ(ancilla[n-2], q[n-1]);}
}&

\specialcell{
\texttt{~~~~Z | q[-1]}
}

\\\midrule

\textbf{5}&

\specialcell{
\texttt{// Undo the local registers}\\
\texttt{for(j=n-2; j>0; j--)}\\
\texttt{~~~~CCNOT(ancilla[j-1], q[j+1], ancilla[j]);}\\
\texttt{CCNOT(q[1], q[0], ancilla[0]);}
}&

\specialcell{
\texttt{\# ProjectQ automatically}\\
\texttt{\# uncomputes control}
}

\\\midrule

\textbf{6}&

\specialcell{
\texttt{// Restore q}\\
\texttt{for(j=0; j<n; j++) X(q[j]);}\\
\texttt{for(j=0; j<n; j++) H(q[j]);}
}&

\specialcell{
\texttt{Uncompute(eng)}
}

\\\bottomrule
\end{tabularx}
\label{fig:grover_code}
\end{table*}

\section{QC program debugging across algorithm primitives}
\label{sec:algorithms}

This section shifts focus away from the Shor's algorithm case study and presents two additional debugging case studies.
The goal is to understand whether the debugging techniques for Shor's algorithm generally apply to other classes of algorithms.

In the Shor's case study, we argued how the structure of the algorithm code guides the placement of assertions.
Our methodology for debugging the algorithm was to bring up the subroutines from unit tests to full integration tests.
We used assertions to check for preconditions, intermediate states, and postconditions of subroutines.
Furthermore the code patterns of how subroutines are composed further guided what assertions to use.
A natural question is whether that rigorous methodology is helpful for debugging other algorithms.

Many different quantum algorithms rely on a handful of QC algorithm primitives to get speedups relative to classical algorithms~\cite{lanl_implementations, montanaro2016quantum, mosca2009quantum}.
These algorithm primitives are akin to algorithm kernels in the context of classical algorithms.
Each algorithm type has distinct pitfalls and features that lead to distinct bugs and possible defenses.

This section covers two more algorithms that use completely different algorithm primitives.
The first is Grover's database search algorithm based on the amplitude amplification primitive.
The second is a quantum chemistry problem that uses quantum operations to simulate a physical system.
This represents a broad selection of different quantum algorithm primitives.

While we have not covered in this paper some algorithm primitives (such as adiabatic algorithms, approximate optimization algorithms, and much less prominent primitives such as quantum random walks), the three areas we have covered represent the most important and well-studied algorithm classes.
\subsection{Case study: Grover's database search}

% The key point spanning both ideas here is you need some guidance on how entanglement should ebb and flow
% entanglement analysis of grovers (done)

This section uses the Grover's benchmark to discuss how language syntax support for reversible computation and controlled operations guides placement of assertions.

\subsubsection{Language support for placement of entanglement assertions}

Higher-level quantum programming language features can help \emph{automatically} place \texttt{assert\_entangled} and \texttt{assert\_product} assertions.
We concentrate on the placement of these two assertion types because they are assertions on the relationship between two or more quantum variables.
As such they are powerful debugging tools,
but they also need the most programmer insight to correctly place them.

As we discussed in Sections~\ref{sec:recursion} and~\ref{sec:mirroring}, entanglement assertions are closely related to the quantum program patterns of recursion and mirroring.
In the Scaffold language,
these patterns are not explicitly captured by the C-style syntax,
but in higher-level quantum programming languages, such as ProjectQ~\cite{Steiger2018projectqopensource} and Q\#~\cite{q_sharp}, these patterns are essential to the language design.
With these language features,
the placement of entanglement assertions becomes as natural as placing precondition and postcondition assertions.

\subsubsection{The Grover's algorithm for database search}

The Grover's search algorithm finds an entry that matches search criteria, among an input data set of size $N$, with a time cost on the order of $\sqrt{N}$.
That represents a polynomial speedup relative to the linear time cost in a classical computer~\cite{grover2001schrodinger}.
% Quantum Grover's algorithm requires O(root(N)) queries
% In the case of Grover’s algorithm, T ~= sqrt(N), where N are the number of database entries
% What is the classical algorithm cost? Lower bound Omega(N); maybe you can do N/2 or some fancy trick, but ultimately still Omega(N).
% linear algorithm

The Grover's algorithm comprises three parts.
First, the input qubits representing the indices of the matching entries are put in a state of superposition, akin to querying all entries at once.
A superposition assertion (Section~\ref{sec:preconditions}) helps certify that this algorithm precondition is satisfied.
% Creation of a configuration where amplitude of $2^n$ basic states are equal
% Walsh-Hadamard transformation operation
Second, the queries are put through a subroutine that checks for the search criteria.
% square root in GF(2)
In our case study, our criteria is to find the square root of a number in a Galois field of two elements, a simple abstract algebra setting.
% superposition, oracle, phase kickback based
% Selective rotation
% in many cases, you have to build an oracle out of a classical function
% Oracle has to correctly implement classical function
% this needs to be done automatically
% Grover's search is more like an extension of Bernstein-Vazirani (and because it's the same circuit, the Deutsch-Jozsa) algorithms
% Grover's oracle is doing phase kickback
Finally in the critical third step, the \emph{amplitude amplification} algorithm primitive amplifies the index that matches the criteria while damping out those that do not.

\subsubsection{Entanglement program patterns in the amplitude amplification subroutine}

Table~\ref{fig:grover_code} shows the reversible computation and controlled operations program patterns coded
in two quantum programming languages Scaffold~\cite{scaffcc} and ProjectQ~\cite{Steiger2018projectqopensource}.
The ProjectQ language has syntax support for reversible computation that automatically mirrors and inverts sequences of operations.
% For example, in Table~\ref{fig:grover_code}, the operations in rows 2 and 3 are respectively mirrored and undone in rows 6 and 5.
Likewise, syntax support for controlled operations automatically allocates the ancillary qubits needed for controlled operations.
% For example in Table~\ref{fig:grover_code},
% rows 3 and 5 are just computing the intersection of qubits \texttt{q},
% in order to realize the controlled rotation operation in row 4.

These higher-level language support for these patterns allows automatic placement of assertions:
the controlled operation statement in rows 3 through 5 indicates that register q should be entangled in row 4,
so it would be the right place to place an entanglement assertion.
Furthermore the compute-uncompute pattern in Rows 2 and 6 hints at a product state assertion at the end of uncomputation.

% Grover's operator is HX cZ XH... what is this significance?

% Grover's progress
% Number of target items in the search database
% Size of database
% avoid overshooting
% verification that number of database entries are the number you expect
% While on the other hand Grover's search algorithm should return the right result after an exact number of iterations.

% Using the knowledge of how operations are paired up in this reversible computing pattern, we can then use static analysis to ensure ancillary qubits are properly freed at the end of each subroutine~\cite{scaffcc}.
% Static analysis for entanglement is not easy because it would appear entanglement just spreads

% Furthermore, the controlled operations syntax works well in combination with the reversible computation syntax, as reversible operations inside of controlled subroutines cancel each other out, leaving much fewer operations that actually need control qubits~\cite{software_methodology}.
% Control(Compute, action, uncompute) is same as Compute, control(action), uncompute
% These simplifications have an impact on efficient simulation of quantum programs, which we discuss in Section~\ref{sec:simulation}.
% N-bit Toffoli -> CCNOTs -> decomposition has T gates
\begin{table*}[ht!]
\centering
\caption{QC calculated energy for $\mathrm{H_2}$ (bond length = 73.48 pm) for different electron assignments.}
\label{fig:hydrogen}
\small
\begin{tabular}{r|p{.75cm}p{.75cm}|p{.75cm}p{.75cm}|l}
\toprule
&\multicolumn{4}{c|}{\textbf{Electron assignments}}&\multirow{3}{3cm}{\textbf{QC calculated energy (relative)}}\\
&\multicolumn{2}{c|}{Bonding}&\multicolumn{2}{c|}{Antibonding}&\\
&$\uparrow$&$\downarrow$&$\uparrow$&$\downarrow$&\\
\midrule
\nth{3} excited state (E3)&0&0&1&1&-0.164\\
\midrule
\multirow{2}{*}{\nth{2} excited state (E2)} &0&1&1&0&\multirow{2}{*}{-0.217}\\
&1&0&0&1&\\
\midrule
\multirow{2}{*}{\nth{1} excited state (E1)} &0&1&0&1&\multirow{2}{*}{-0.244}\\
&1&0&1&0&\\
\midrule
Ground state (G)&1&1&0&0&-0.295\\
\bottomrule
\end{tabular}
\label{tab:hydrogen}
\end{table*}

\subsection{Case study: Quantum chemistry}
Next, we discuss our experience building up and debugging a simple quantum chemistry program.
Quantum chemistry problems entail finding properties of molecules from theoretical first principles~\cite{qchem, qchem_nsf}.
Researchers anticipate these will be the first applications for QC due to the relatively few number of qubits they need to surpass classical computer algorithms.
Debugging these problems is distinctively challenging, due to the importance of getting a large number of classical input parameters all correct, and because of the dearth of physically meaningful intermediate states we can check in the course of algorithm execution.

\subsubsection{Classical input parameters}
\label{sec:classical_bug}
A key part of quantum chemistry programs is in correctly building up a \emph{Hamiltonian} subroutine that simulates inter-electron forces.
The procedure for doing this was laid out in detail by Whitfield~\cite{whitfield}.
We followed this procedure to create a subroutine for simulating the hydrogen molecule, but we needed additional cross-validation from several other sources to get a bug-free subroutine~\cite{gate_count}.
These resources include raw chemistry data found in open source repositories for the LIQUi|> framework\footnote{\url{https://github.com/StationQ/Liquid/blob/master/Samples/h2_sto3g_4.dat}
}.
% Hartree-Fock parameters for Hamiltonian simulation
% The inputs to these algorithms include assumptions about molecules such as how far atoms are apart, and precomputed parameters that describe electron interactions.
The final parameters for actual operations on qubits were validated against a follow-up paper~\cite{bk_transform} and an implementation in the QISKit framework\footnote{\url{https://github.com/Qiskit/aqua/blob/master/test/H2-0.735.json}}.
Because the procedure for preparing these quantum chemistry models involves many steps and needs domain expertise,
arguably this step in preparing classical input parameters is the hardest aspect to debug.
% software packages such as OpenFermion now automate this process~\cite{openfermion}.
% Nonetheless, there is room for improvement in standardizing input data formats to eliminate bugs in this process.

Once the Hamiltonian subroutine is built, we can use the model in a variety of quantum algorithms spanning different primitives.
These include phase estimation (an application of quantum Fourier transforms)~\cite{qpe}, variational quantum eigensolvers~\cite{vqe}, and adiabatic algorithms~\cite{adiabatic}.
In this case study, we use iterative phase estimation to find the ground state energy of our $\mathrm{H_2}$ model, validating results published by Lanyon~\cite{lanyon}.

\subsubsection{Assertions on quantum intial values and final states}
\label{sec:allocation}

The correct preparation of qubit initial values stands out as another challenging aspect of debugging quantum chemistry QC programs.
Incorrect initial values would cause the program to find solutions to different problems altogether.
In this quantum chemistry problem, the initial values control the locations of the two electrons in $\mathrm{H_2}$.
As shown in Table~\ref{tab:hydrogen},
precondition assertions check the qubit assignment for finding the ground energy of $\mathrm{H_2}$,
while other assignments lead to results for other energy levels.

The symmetry of $\mathrm{H_2}$ allows us to perform a sanity check,
to make sure the Hamiltonian and the iterative phase estimation subroutines are working correctly.
Though there are six ways to assign two electrons to four locations,
there are in fact only four distinct energy levels, as shown in the experimental data (Table~\ref{tab:hydrogen}).
Postcondition assertions are useful for checking that the two different ways to obtain E1 (and E2) give the same energy levels.
These assertions validate that the model correctly preserves symmetry.

\subsubsection{Assertions on intermediate algorithm progress}
\label{sec:progress}

Unlike the other two case studies in this paper, the debugging process for the quantum chemistry benchmark is coarse-grained.
That is because the Hamiltonian subroutine is a monolithic block of code whose components do not have obvious expected outputs---its components represent pair-wise electron interactions, and do not have inherent physical meaning.
So how do we debug this program? The preconditions in the last subsection make sure the inputs to the algorithm are correct; the other observable state we have for debugging is to check the behavior of the algorithm as a whole.

In this quantum chemistry program, we can check for two types of overall algorithm behavior.
One is the solution should converge to a steady value as finer Trotter time steps (a kind of numerical approximation) are chosen; a lack of this type of convergence indicates a bug in the Hamiltonian subroutine.
The other algorithm behavior is when we vary the precision of the phase estimation algorithm,
the most significant bits of the measurement output sequences should be the same---in other words, rounding the output of a high-precision experiment should yield the same output as a lower-precision experiment.
a lack of this convergence indicates a bug in the iterative phase estimation subroutine.
These checks for expected algorithm progress also apply to other algorithms.

% Phase estimation estimates phi in an operator eigenvalue exp(2*pi*I*phi)
% Measuring in multiple bases to get phase estimation
% QPE error margin of QPE

% Minimum gap condition in adiabatic optimization
% Can energy state ordering be reordered during adiabatic evolution?

\section{Related work}
\label{sec:related}
% Several related approaches have been proposed in the area of assertions, debugging, and verifying quantum programs.
Approaches to writing correct quantum programs range from formal methods to less-formal pragmatic methods, much like in classical programming.
Most of the prior work in quantum program correctness has been in formal methods--e.g., using theorem provers and type checking to verify programs correctly match algorithm specifications~\cite{ying_hoare,invariants,quantitative_robustness,rand2018formally,qwire}.
Such verification techniques consider the correctness problem from a top-down perspective.
While useful, formal methods should not be the only approach to correct programs; more traditional debugging strategies are also useful.
This work considers the possibility of using pragmatic assertion checks to build code bottom-up from exhaustive unit tests up through integration testing.

Quantum program assertions exist in several quantum programming languages,
though not in the same capacity as the statistical assertions presented in this work.
First, the Quipper language~\cite{quipper} features \emph{assertive termination}, which allows the programmer to annotate known program state, in order to drive compiler optimizations.
Their use relies on the programmer to write correct code and assertions, and cannot be used as postcondition checks.
Second, the Q\# programming language~\cite{q_sharp} allows programmers to write assertions on classical, integer states.
These assertions are then checked during the simulation of the quantum programs.
This work extends that set of assertions with assertions on superposition and entanglement states.

The set of quantum states considered in these assertions,
namely classical, superposition, and entangled states,
are a subset of possible quantum states.
In the general case we need quantum phase estimation, quantum state tomography, and quantum process tomography to be able to examine general quantum states~\cite{nielsen_chuang}.
However, those processes are extremely costly and cannot be used as efficient assertion checks.

% Quantum information theory provides an important set of metrics such as relative entropy and mutual information~\cite{nielsen_chuang},
% which quantify the extent to which quantum states are in superposition or entanglement.

\section{Conclusion}

For the first time, we have access to program benchmarks for several major areas of quantum algorithms,
along with input datasets and outputs that are detailed enough to permit detailed debugging and cross-validation.
Using our experience in building up and debugging these programs, we presented in this paper a strategy for deploying and checking quantum program assertions based on statistical tests.
Drawing on the structure of quantum programs, we point to where and how program bugs may arise, and point to how the presented assertions can catch them.

% defense strategies that facilitate writing bug-free QC code, summarized in Table~\ref{tab:apply_matrix}.
% Many of these ideas, such as refactoring code, data types, and assertions, are commonplace and even obvious in classical programming languages.
% But only recently has QC infrastructure become mature enough to need and support these ideas.
% Successful transplantation of these ideas from classical languages to QC languages can pave the way towards correct and useful quantum programs.

\begin{acks}
This work is funded in part by EPiQC, an NSF Expedition in Computing,
under grant 1730082.
\end{acks}

\bibliographystyle{ACM-Reference-Format}
\bibliography{qdb_isca}

\end{document}